\documentclass[a4paper,11pt]{article}
\usepackage{pos}
\usepackage{graphicx}
\usepackage{float}
\usepackage{wrapfig}
\usepackage{hyperref}

\newcommand{\barechiralcondinlattice}{\left \langle \Bar{\psi}\,\psi \right \rangle}
\newcommand{\barechiralcond}{M_\ell}
\newcommand{\barechiralsusc}{\chi_\ell}

\title{Towards a parameter-free determination of critical exponents and
chiral phase transition temperature in QCD}
\ShortTitle{Critical exponents and chiral phase transition temperature in QCD}

\author*[a]{Sabarnya Mitra}
\author[a]{Frithjof Karsch}
\affiliation[a]{Fak{\"u}ltat F{\"u}r Physik, Universit{\"a}t Bielefeld, D-33615 Bielefeld, Germany}
\author[b]{Sipaz Sharma}
\affiliation[b]{Physik Department, Technische Universit\"at M\"unchen, D-85748 Garching~b.~M\"unchen, Germany}

\emailAdd{smitra@physik.uni-bielefeld.de}

\abstract
{
In order to quantify 
the universal properties of the
chiral phase transition
in (2+1)-flavor QCD,
we make use of
an improved, renormalized order parameter
for chiral symmetry breaking which is obtained as a suitable difference of the $2$-flavor light quark chiral condensate and its corresponding light quark susceptibility. Having no additive ultraviolet as well as multiplicative logarithmic divergences, we use  
ratios of this order parameter constructed
from its values for two different light quark masses. We show that this facilitates determining in a parameter-independent manner, the chiral phase transition temperature $T_c$ and the associated critical exponent $\delta$ which, for sufficiently small values of the light quark masses, controls the quark mass dependence of the order parameter at $T_c$.  
We present first results of these
calculations from our numerical analysis
performed with staggered fermions
on $N_\tau=8$ lattices.  
}

\FullConference{The 41st International Symposium on Lattice Field Theory (LATTICE2024)\\
 28 July - 3 August 2024\\
Liverpool, UK\\}

\begin{document}
\maketitle

\section{Introduction}

One of the outstanding questions in the study of the phase diagram of 
strong-interaction matter is whether or not, the effective restoration of 
the axial anomaly has an influence on the universal properties of the Quantum Chromodynamics (QCD) chiral phase transition. It has been argued \cite{Pisarski:1983ms} that a first order
phase transition may occur in QCD with two
light quarks, if the axial anomaly is effectively restored in the limit of 
vanishing up and down quark masses. Moreover
in the case of three light quark flavors, the 
chiral phase transition should be first order
irrespective of the temperature dependence 
of the chiral anomaly.

More recently, it has been shown that in the case of two light flavors, it still would be
possible to have a second order phase 
transition belonging to the 3-dimensional $U(2)\times U(2)$ universality class even,
when the axial $U(1)_A$ symmetry gets 
restored in the chiral limit \cite{Pelissetto:2013hqa}.
This is in contrast to universal critical
behavior in the 3-dimensional, $O(4)$ universality class which, one would expect
to control the thermodynamics of QCD
in the limit of vanishing two light quark masses
(2- or (2+1)-flavor QCD), if the axial anomaly does
not get restored. Here we aim at a
first principle lattice QCD analysis of 
universal critical behavior in QCD with 
two light, degenerate up and down quarks and 
a heavier strange quark mass tuned to its physical value. This does build on earlier
studies of the magnetic equation of state
in (2+1)-flavor QCD using staggered
fermion discretization schemes
\cite{Ejiri:2009ac,Bazavov:2011nk,Ding:2024sux}.

Even when the chiral phase transition in QCD belongs to the
3-dimensional, $O(4)$ universality class, it is an open question to which extent this 
is reflected in the thermodynamic properties
of QCD with its physical spectrum of light
and strange quark masses. In order to
address this question, a detailed analysis is needed
regarding the temperature and quark 
mass dependence of e.g. the chiral order parameter and its 
susceptibility in a narrow temperature interval in the vicinity of the chiral
phase transition temperature. This will allow us to disentangle the universal critical behavior, which eventually shows up as divergence of higher order derivatives of the free energy, and the non-critical
universal correction-to-scaling
as well as regular
contributions to thermodynamic quantities. 
It will also allow us to quantify the size of the scaling regions relevant for 
different physical observables 
\cite{Kotov:2021rah,Braun:2023qak}
and thus, will
provide crucial input to the interpretation of
experimentally measured observables, e.g.
higher order cumulants of net-baryon number
fluctuations and the correlations between
different conserved charges \cite{Bazavov:2020bjn}.

\section{Scaling behaviour and scaling functions} 
 
\subsection{Theoretical background}
We give here, some background for our study of the universal structure of thermodynamic observables in the vicinity of the chiral phase transition in $(2+1)$-flavor QCD. The relevant quantities of interest are the $2$-flavor light quark chiral condensate, $\barechiralcond$ and the corresponding chiral susceptibility $\barechiralsusc$, which is obtained as the derivative of the light quark chiral condensate with respect to the light quark mass $m_\ell$, 
  \begin{equation}
      \barechiralcond = \frac{m_s}{f_K^4} \barechiralcondinlattice, \hspace{1cm}
      \barechiralsusc = m_s \,\frac{\partial \barechiralcond}{\partial m_\ell} \; ,
      \label{basics}
  \end{equation}
  where factors of the strange quark mass, $m_s$, are introduced as multiplicative renormalization factors and the kaon decay constant, $f_K$, is used to obtain dimensionless quantities. 
The chiral condensate $M_\ell$ still requires
additive renormalization in order to arrive at
an order parameter for chiral symmetry breaking
that is well defined in the continuum limit. 
The ultraviolet divergent contributions to $M_\ell$
  may be eliminated by subtracting a suitable
  fraction of $\chi_\ell$. We therefore introduce 
  as an order parameter for chiral symmetry 
  breaking, the difference 
 \begin{equation}
     M(T,H) = \barechiralcond (T,H) - H\,\barechiralsusc(T,H)\; ,
     \label{eq:M}
 \end{equation}
with $H$ given as the ratio of light and strange quark masses, $H\equiv m_\ell/m_s$. 
It is obvious from the definition of $\chi_\ell$ given in Eq.~\ref{basics} that terms linear in
$m_\ell$ are explicitly eliminated
in $M(T.H)$. In the vicinity of the chiral limit this order parameter thus receives only light quark mass  corrections that are proportional to $H^3$.
We also note that 
we treat the strange quark mass just like $f_K$
as an external parameter, that is tuned to its physical value by demanding that the mass of
the pseudoscalar
meson, $\eta_{\bar{s}s}$, is fixed to its physical value. The corresponding line of constant physics
has been introduced in \cite{HotQCD:2014kol}.
This version of a renormalized order parameter
has been introduced in \cite{Unger:2010wcq} and has been used previously to analyze critical behavior in lattice QCD calculations \cite{Kotov:2021rah,Dini:2021hug,Ding:2024sux}. 
In the vicinity of the chiral critical point, $(T,H)=(T_c,0)$,  
the temperature and quark mass dependence of this version of an
order parameter is controlled by a scaling function, which is given by the difference of 
the commonly
introduced order parameter scaling function, $f_G(z)$ and the corresponding susceptibility scaling function, $f_\chi(z)$. Near
the critical point, these scaling functions control the universal critical behavior of the unrenormalized order parameter $M_\ell$ and its susceptibility,
respectively,
  \begin{eqnarray}
      M_\ell(T,H)  &=& h^{\frac{1}{\delta}} \, f_G(z) 
      + M_{\ell,\,sub-lead}(T,H)\; , 
      \\
      \chi_{\ell}(T,H)  &=& \frac{1}{\delta}\,h^{^{\frac{1}{\delta}-1}} f_{\chi}(z) +\chi_{\ell,\,sub-lead}(T,H)
      \; .
      \label{eq:bare M and susc}
  \end{eqnarray}
 Here $\delta$ is the critical exponent,
 controlling the dependence of the order parameter as function of the symmetry 
 breaking parameter $H$ at $T=T_c$, {\it i.e.} $M(T_c,H)= h^{1/\delta}$, with $h=H/h_0$ and $h_0$ denoting
 a non-universal constant. Close to the critical point, the $T$ and $H$ dependence
 of $M$ is controlled by a single scaling variable,
 \begin{equation}
     z = t\,h^{-1/\beta\delta}, \hspace{.6cm} \text{where} \hspace{.4cm} t = \frac{1}{t_0}\,\left[\frac{T}{T_c}-1\right], \hspace{.4cm} h = \frac{H}{h_0}\; ,
     \label{eq:scaling variable}
 \end{equation}
 with $t_0$ denoting another non-universal
 constant.
 In addition to the leading non-analytic (so-called singular) behavior, the order parameter
 receives corrections arising from sub-leading universal corrections-to-scaling as well as analytic, non-universal terms. We denote both of them as sub-leading corrections.
 The leading, singular behavior of the order parameter $M$ is then described by,
 \begin{equation}
     M(T,H)= h^{1/\delta} f_{G\chi}(z) +~{\rm sub-lead.}\; , \;\; {\rm with} \;\; f_{G\chi}(z)= f_{G}(z)-f_{\chi}(z)
     \; .
     \label{MTH}
 \end{equation}
 Using the relation $f_{\chi}(z)= \delta^{-1}\left(f_{G}(z)-zf'_{G}(z)/\beta\right)$, with
 $\beta$ denoting a critical exponent, we have
 \begin{equation}
     f_{G\chi}(z)= \left( 1-\frac{1}{\delta}\right)
     f_{G}(z) +\frac{z}{\beta\delta}\,f'_{G}(z) \; ,
 \end{equation}
 and in particular, $f_{G\chi}(0)=1-1/\delta$.
 
\begin{figure*}[t]
\includegraphics[width=0.47\textwidth]{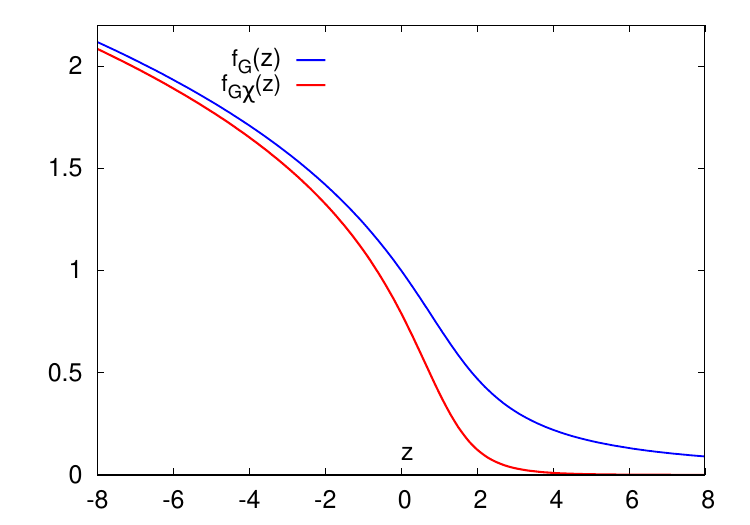}
\includegraphics[width=0.47\textwidth]{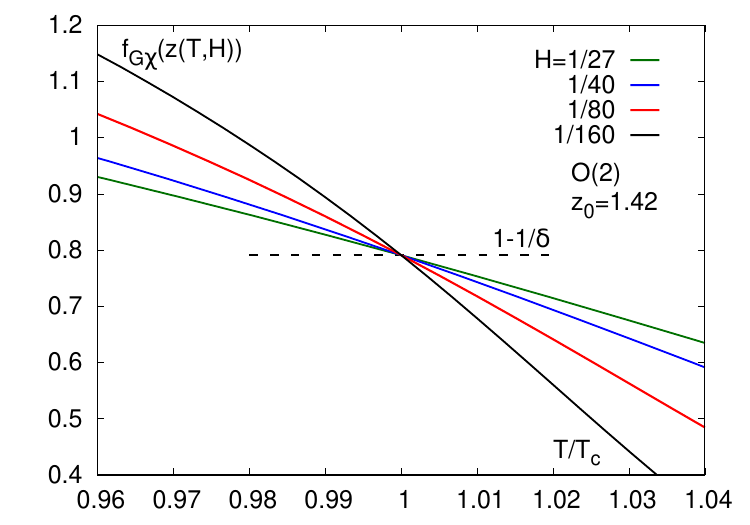}
\caption{{\it Left:} The 3-$d$, $O(2)$ scaling functions $f_G(z)$ and $f_{G\chi}(z)$ as function of scaling variable $z$.
{\it Right:} The 3-$d$, $O(2)$, scaling function $f_{G\chi}(z)$ versus $T/T_c$ for an arbitrary non-universal parameter value, $z_0=1.42$. }
\label{fig:fGfGchi}
\end{figure*}

The universal scaling functions $f_{G}(z)$ and $f_{\chi}(z)$
have been determined for several universality classes. For a recent parametrization, see for 
instance \cite{Karsch:2023pga}. 
These scaling functions differ in detail, but
show qualitatively similar behavior. E.g.
the value of $f_G(z)$ and as such also of 
$f_{G\chi}(z)$ at $z=0$ directly gives the
critical exponent $\delta$ and both scaling
functions have a unique intersection point at
$z=0$ when plotted versus $T$ for different values
of $H$. 
In Fig.~\ref{fig:fGfGchi}~(left), we show the scaling functions $f_G(z)$ and $f_{G\chi}(z)$
 for the 3-dimensional $O(2)$ universality class, which is most relevant
for the current scaling analysis performed by
us at a single finite lattice cut-off corresponding to lattices with temporal extent $N_\tau=8$.
In the right hand part of Fig.~\ref{fig:fGfGchi},
we show $f_{G\chi}(z)$ as function of $T/T_c$ 
for several values of $H$. The choice for the 
latter corresponds to light-to-strange quark
mass ratios used previously also in a determination of the chiral phase transition temperature in $(2+1)$-flavor QCD \cite{HotQCD:2019xnw}. To draw this figure, we 
used for the non-universal scale parameter
$z_0=h_0^{1/\beta\delta}/t_0$, the value $1.42$,
which has been determined in a previous scaling 
analysis of the order parameter $M$ \cite{Ding:2024sux}. However, obviously the qualitative features seen in Fig.~\ref{fig:fGfGchi}~(right) do not depend on this particular choice.
  
 \subsection{Universal order parameter ratios}
 
From Eq.~\ref{MTH}, one obtains the rescaled order 
 parameter
 \begin{equation}
      M(T,H)/H^{1/\delta} = h_0^{-1/\delta}
     f_{G\chi}(z) + {\rm sub-lead.} \; ,
 \end{equation}
which allows to determine the chiral
 phase transition temperature from the 
 unique intersection point, 
once the sub-leading corrections are sufficiently small.
 Its construction, however requires known
 knowledge of the relevant universality class, {\it i.e.} the critical exponent $\delta$.
 To avoid the need for 
this input, we construct
 ratios of the chiral order parameter evaluated
 for $H$-values that differ by a factor $c$, like
 \begin{equation}
     R(T, H, c) = \frac{M(T,cH)}{M(T,H)} \; .
 \label{M-ratio}
 \end{equation}
 
 Obviously also these ratios, evaluated for different values of $H$, have a unique intersection
 point at $T_c$ once the sub-leading corrections become
 negligible. We may use these ratios to obtain
 directly the critical exponent $\delta$ at
 this intersection point. We introduce

\begin{figure}
\begin{center}
\includegraphics[scale=0.75]{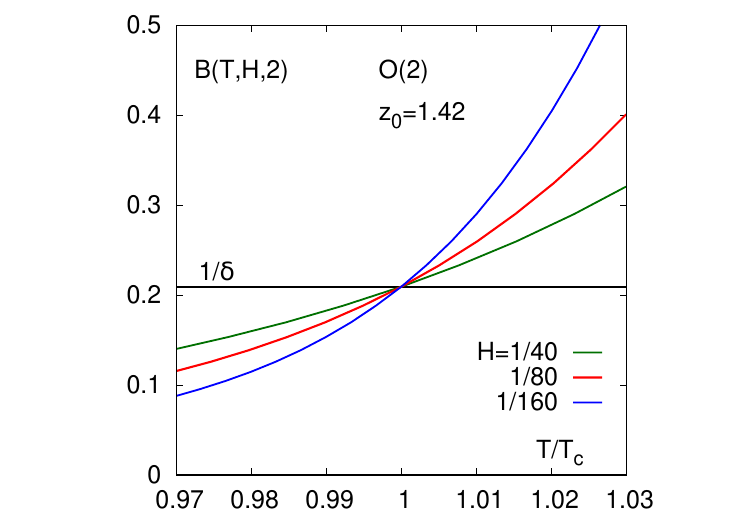}
\caption{Logarithm of the ratio of order parameters, $B(T,H,c)$, as defined in Eq.~\ref{M-ratiolog} versus $T/T_c$. Shown
is the result obtained in the $3$-dimensional $O(2)$ universality class for $c=2$ and the non-universal scale $z_0=1.42$.
}
\label{fig:Mratio}
\end{center}
\end{figure}
 \begin{eqnarray}
     B(T,H,c) &=& \frac{\ln R(T,H,c)}{\ln (c)}\; .
 \label{M-ratiolog}
 \end{eqnarray}
 
 This ratio is shown in Fig.~\ref{fig:Mratio} for the $3$-dimensional $O(2)$ universality class using the scaling functions shown in Fig.~\ref{fig:fGfGchi}.
 In the chiral limit one finds at $T_c$,
 \begin{equation}
     \lim_{H\rightarrow 0} B(T_c,H,c) = \frac{1}{\delta} \; .
 \end{equation}
 In the following we will present first results
 for the ratio $B(T,H,c)$
 obtained in simulations 
 of (2+1)-flavor QCD on lattices with temporal
 extent $N_\tau=8$.

\section{Computational setup}

The preliminary results for the logarithm 
of the order parameter ratio $B(T,H,2)$ discussed
in the next section have been obtained from
lattice QCD calculations in $(2+1)$-flavor QCD
using the HISQ action and the $\mathcal{O}(a^2)$ improved Symanzik gauge action. Our computational setup is identical to that used previously in the 
determination of the chiral phase transition from
simulations with smaller than physical light quark masses \cite{HotQCD:2019xnw} as well as the recent determination of the chiral phase transition temperature at non-vanishing chemical potential \cite{Ding:2024sux}. These calculations have been
performed on lattice of size $N_\sigma^3\times 8$
with light to strange quark mass ratios $1/160\le H\le 1/20$ corresponding to pion masses $55~{\rm MeV}\le m_\pi\le 160~{\rm MeV}$. The strange
quark mass has been fixed to its physical value, demanding that the strange $\eta$ meson mass defined in terms
of kaon and pion masses, $M_{\bar{s}s}=\sqrt{2 m_K^2-m_\pi^2}$,
is kept fixed \cite{HotQCD:2014kol}.

In previous work the spatial lattice extent $N_\sigma$ has been
increased with decreasing $m_\pi$, {\it i.e.}
$4\le N_\sigma \le 7$, insuring that the 
inverse of the pion correlation length in
units of the spatial lattice extent stays 
approximately constant, $m_\pi L \simeq (3-4)$.

For our analysis of the rescaled order parameter and the order parameter ratios
we used the set of data for $H=1/27, 1/40, 1/80$ and $1/160$ at several values of the 
temperature, tabulated in \cite{Ding:2024sux}. Data for $H=1/20$
are taken from \cite{Ding:2024sux}. These data are generally 
taken at temperature values separated by
$(2-3)$~MeV. In order to arrive at precision
determination of the unique crossing point 
in the rescaled order parameter, small separations of the $T$-values is needed.
Further, more smaller separations in the set
of $H$ values is needed to control the
quark mass dependence of the chiral order
parameter. We started on this program by
adding new data at $T=144.05$~MeV and at
additional $H$-values, $H=1/32.2$ and $1/16.1$.
The current statistics at these new sets of
$(T,H)$ values is about $\mathcal{O}(20K)$ gauge configurations, separated by 5 and 10 RHMC time units, respectively.

\section{Results} 

\begin{figure*}
\hspace{-1.7cm}
\includegraphics[width=0.70\textwidth]{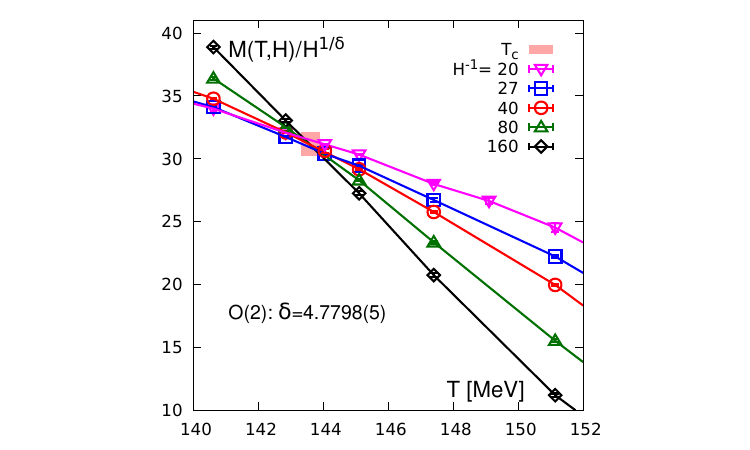}
\hspace{-3.4cm}
\includegraphics[width=0.70\textwidth]{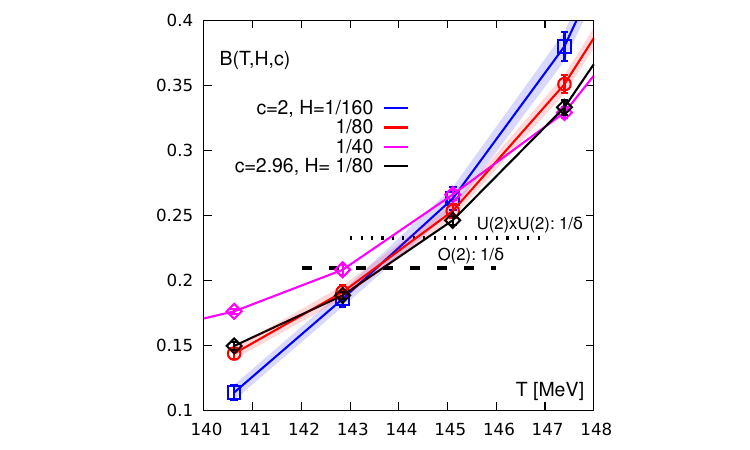}
\caption{{\it Left:} Rescaled order parameter versus temperature for several values of $H$.
The shaded rectangle corresponds 
to the values of $(T_c,h_0^{-1/\delta})$ determined
previously from scaling fits 
\cite{Ding:2024sux}.
{\it Right:} The logarithm of ratios of the 
chiral order parameter versus temperature as introduce in Eq.~\ref{M-ratiolog}. The dashed and dotted lines give the values for $1/\delta$ in the $O(2)$ and $U(2)\times U(2)$ \cite{Pelissetto:2013hqa} universality classes, respectively.
}
\label{fig:ordera}
\end{figure*}

In Fig.~\ref{fig:ordera}~(left), we show results for the rescaled order parameter
$M/H^{1/\delta}$ taken from \cite{Ding:2024sux}. Here we used for 
the critical exponent $\delta$, the value in
the $3$-dimensional $O(2)$ universality class, $\delta=4.7798$. The slope of the rescaled order parameter rises with decreasing $H$ and the existence of an intersection 
point is clearly visible. In fact, the location of this 
intersection point, is seen to agree well 
with the previous determination of $T_c$ 
{\it i.e.} $T_c^{N_\tau=8}=143.7(2)$~MeV
\cite{Ding:2024sux}.
Also the value of $M/H^{1/\delta}$ at this intersection point,
\begin{equation}
    M(T_c,H)/H^{1/\delta}=h_0^{-1/\delta} (1-1/\delta)\;
\end{equation}
agrees well with the previously determined value, $h_0^{-1/\delta}=39.2(4)$ \cite{Ding:2024sux}, which has
been obtained from fits to the rescaled order parameter in the 
temperature interval $T\in [140\,{\rm MeV}:148\,{\rm MeV}]$.
We note however, again that this 
analysis starts with assuming a 
second order critical point in the $3$-dimensional 
$O(2)$ universality. In order to 
avoid such an apriori input into the
analysis, we may turn to an analysis
of ratios of the order parameter $M$
as introduced in Eq.~\ref{M-ratiolog}.

In Fig.~\ref{fig:ordera}~(right) we show
result for the logarithm of ratios of the 
chiral order parameter versus temperature for several sets of 
$H$-value, $(H_1,H_2)$ with $H_2=c H_1$ and $H_1=1/40, 1/80, 1/160$. 
While results for $B(T,H,c)$ with
$cH\le 1/40$ give a unique intersection 
point that is in agreement with a
critical exponent $\delta$ in the 
$O(2)$ universality class, this is not the case once larger values of $cH$
are involved in the order parameter ratios entering the definition $B(T,H,c)$. In Fig.~\ref{fig:ordera}~(right),
this is the case for $B(T,1/80,80/27)$, and also for $B(T,1/40,2)$.
This suggests that the quark ratio
$H=1/27$, corresponding to the physical value of light and strange quark masses and larger than $1/27$ ratios are too large and clearly
not in the scaling region where contributions from regular terms
can be neglected. Clearly, more detailed information on the $H$-dependence of the rescaled order parameter, $M/H^{1/\delta}$, is needed
to quantify the parameter range in which, a unique crossing point can be established that would then also allow for a clear-cut distinction between different universality classes.

\section{Conclusion and Outlook}

We have shown that, the rescaled order parameter evaluated as function of $T$
for different values of the light-to-strange quark mass ratio $H=m_\ell/m_s$ exhibits the feature
of having a unique intersection point, if the light quark mass or respectively
$H$, is small enough so that contributions from sub-leading terms can be neglected. Taking ratios of order parameters evaluated for different values of $H$ allows for a determination of the critical exponent 
$\delta$ without making apriori assumption about the underlying universality class.
However in order to arrive at the point, where these calculations reach a sufficient accuracy to allow for quantitative determination of $\delta$
thereby enabling a clear distinction between universality classes relevant 
for the analysis of the chiral phase transition, one needs to collect more 
information on the $H$ dependence of 
the rescaled order parameter close to the chiral limit. Furthermore, a more detailed analysis of finite volume and cut-off effects will be needed that were not subject of the discussion presented here. In fact, this is work in progress.

\section*{Acknowledgments}
This work was supported by the Deutsche Forschungsgemeinschaft
(DFG, German Research Foundation) Proj. No. 315477589-TRR 211. Numerical computations
have been performed on GPU-clusters at Bielefeld University.

\end{document}